\documentclass[conference,a4paper]{IEEEtran}

\usepackage[letterpaper, left=1in, right=1in, bottom=1in, top=0.75in]{geometry}
\usepackage{amsthm}
\usepackage{amsmath}
\usepackage{amssymb}
\usepackage{graphicx}
\usepackage{esint}

\makeatletter
\theoremstyle{plain}

\usepackage{amsfonts}
\usepackage{cite}
\usepackage{balance}
\usepackage{caption}
\usepackage{subcaption}
\usepackage{epstopdf}

\IEEEoverridecommandlockouts

\providecommand{\theoremname}{Theorem}

\makeatother

\providecommand{\theoremname}{Theorem}

\begin{document}

\author{\IEEEauthorblockN{Evangelos N. Papasotiriou\IEEEauthorrefmark{1}, Alexandros-Apostolos A. Boulogeorgos\IEEEauthorrefmark{1}, and  Angeliki Alexiou\IEEEauthorrefmark{1}
} \IEEEauthorblockA{\IEEEauthorrefmark{1}{\footnotesize{}{}{{{{{{{Department of Digital Systems, University of Piraeus, Piraeus 18534, Greece. }}}}}}}} 
\\
E-mails:  \{vangpapasot, alexiou\}@unipi.gr, al.boulogeorgos@ieee.org 
} 
}
%
\title{Performance evaluation of THz wireless systems under the joint impact of \\ misalignment fading and phase noise}
%

\maketitle


\begin{abstract}
In this paper, we investigate the joint impact of misalignment fading and local oscillator (LO) phase noise (PHN) in multi-carrier terahertz (THz) wireless systems. 
In more detail, after establishing a suitable system model that takes into account the particularities of the THz channel, 
as well as the transceivers characteristics, 
we present 
simulation results that quantify the joint impact of misalignment fading and PHN in terms of average signal-to-interference-plus-noise-ratio (SINR) and outage probability (OP).
\end{abstract}

\IEEEpeerreviewmaketitle

\vspace{-0.3 cm}
\section{Introduction}
\vspace{-0.2 cm}
Spectrum limitations have motivated researchers to look at the THz frequencies for beyond the 5G wireless networks. THz bands have an advantage of ultra high bandwidth availability; hence extremely high data rates could be supported~\cite{ref1_VTC}. 
However, THz systems suffer from channel attenuation significantly limiting the transmission range, hence making necessary the use of highly directive antennas at both the base station (BS) and user equipment (UE)~\cite{A:Analytical_Performance_Assessment_of_THz_Wireless_Systems}. Furthermore, the transceivers hardware imperfections mainly caused by the LO imperfections~\cite{ref14_Al_hard_imperf} severely degrade the performance of wireless radio frequency (RF) systems. In this sense, to the best of the authors knowledge, there is no study on the joint impact of misalignment fading and LO PHN in the performance of THz wireless systems. 
%
%

\vspace{-0.2 cm}
\section{System  Model \& Performance metrics}\label{S:SM}
\vspace{-0.2 cm}
\underline{System model}: We consider a downlink scenario in a multi-carrier wireless THz system with a single BS and a UE, equipped with $N_b$ and $N_u$ antennas, respectively. The wideband RF signal of bandwidth $W$ is assumed to be divided to $K\in\mathbb{Z}$ narrowband carriers and each carrier $k\in\lbrace -\frac{K}{2},...,-1,1,...\frac{K}{2}\rbrace$ has a bandwidth of $W_{ch}=W_{sb}+W_{gb}$, where $W_{sb}$ denotes the transmitted signal bandwidth and $W_{gb}$ is the guard band bandwidth. 
Both the BS and UE are assumed to perform analog beamforming and that the UE suffers from LO PHN, a practical assumption, due to the low-cost demand and the low-computational capabilities of the UE~\cite{A:IQSC}, while the BS has a perfect RF chain. 
The baseband equivalent received signal at the carrier $k$, assuming that the beamforming directions are orthogonal, can be obtained~\cite{ref14_Al_hard_imperf}
\vspace{-0.2cm}
\begin{align}
y_{k}=h_{k}s_k+\psi_k+w_k,
\label{received_signal}
\vspace{-0.2cm}
\end{align}
where $s_k$ and $w_k$ are the emitted signal  and the complex zero-mean additive white Gaussian noise (AWGN) at carrier $k$,
while $h_k$ 
stands for MIMO channel coefficient at the $k-$th carrier, which can be expressed as $h_k=h_lh_ph_f$, where $h_l$,$h_p$,$h_f$ respectively model the deterministic path gain, the misalignment fading 
and the stochastic path gain. The deterministic path gain coefficient can be expressed as $h_l=h_{fl}h_{al}$, where $h_{fl}$ models the propagation gain according to the Friis equation, while $h_{al}$ stands for the molecular absorption gain and can be evaluated as in~\cite{EuCAP2018_cr_ver7}.
 For the pointing errors, the probability density function (PDF) of the channel coefficient, $h_p$, can be obtained as in~\cite{A:Analytical_Performance_Assessment_of_THz_Wireless_Systems}.
The fading coefficient $|h_f|$ is modeled as a Nakagami-$m$ process.
Finally, $\psi_k$ denotes the LO PHN vector and can be obtained as, 
$\psi_k = \theta_{k-1} h_{k-1}  \gamma_{k-1}  s_{k-1} + \theta_{k+1} h_{k+1} \gamma_{k+1} s_{k+1},$
 with $\gamma_k=\exp\left({j\phi_k}\right)$ and $\phi_k=\sum_{m=1}^{n}\phi_k\left(m-1\right)+\epsilon\left(n\right)$, where $\phi_k(n)$ is the $n-$th sample of $\phi_k$ and $\epsilon\left(n\right)$ is  a zero mean real Gaussian variable with variance, $\sigma_{\epsilon}^{2}=4\pi\beta / W$, where $\beta$ is the $3\text{ }\rm{dB}$ bandwidth of the LO~process. Moreover, \\$\theta_k=\left\{\begin{array}{l} 1,\text{ }k\in\left[-K/2 - 1, K/2 +1\right] \\ 0\text{, } \text{otherwise} \end{array}\right.$.

\underline{Performance metrics}:
 Based on~\eqref{received_signal} the instantaneous SINR can be obtained as
$\rho = \left|h_k\right|^2 P_{k}/(\sigma_{\psi_k}^2 + N_o),$ 
where $P_k$ and $N_o$ respectively stand for the transmitted signal and noise power, while $\sigma_{\psi_k}^2$ is the variance of the ICI term, which, for a given channel realization, can be evaluated~as  
$\sigma_{\left.\psi\right|\left\{ h_{k},\theta_{k}\right\} }^{2} 
=\theta_{k-1}A_{k-1}\left|h_{k-1}\right|^{2} P_{k-1}
+\theta_{k+1}A_{k+1}\left|h_{k+1}\right|^{2} P_{k+1}$
,where $P_{k\pm1}$ are the transmission powers at carriers $k \pm 1$. Moreover $A_{k \pm 1}$ denote the ICI form carriers $k \pm1 $ and can be obtained as in~\cite{ref14_Al_hard_imperf}.
The OP is defined as the probability that the received signal is below a predetermined threshold, $\gamma_{th}$, that is connected with the transmission scheme spectral efficiency, $r$, through $\gamma_{th} = 2^{r-1}$.

\vspace{-0.2 cm}
\section{Results \& Conclusions}\label{S:RD}
\vspace{-0.2 cm}
We consider that both the BS and UE are equipped with high directive antennas with transmission and reception gains both equal $55\text{ }\mathrm{dBi}$~\cite{A:Analytical_Performance_Assessment_of_THz_Wireless_Systems}, also we assume 
 $m=4$
 and $K=10$ carriers of the $2\text{ }\mathrm{GHz}$ bandwidth each  and total guard bandwidth of $5\text{ }\mathrm{MHz}$. Finally, we set the central frequency to $335\text{ }\mathrm{GHz}$.

\begin{figure}
\centering\includegraphics[width=0.72\linewidth,trim=0 0 0 0,clip=false]{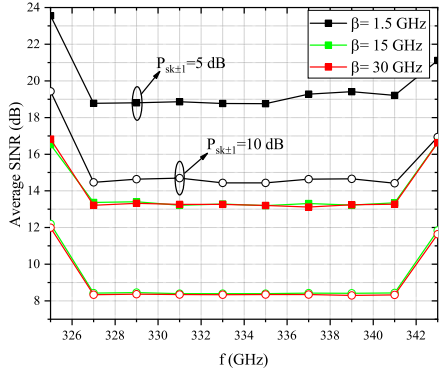}
\caption{Average SINR vs $f_k$ for different values of $\beta$, $P_{k\pm1}$, and  $d=10\text{ }m$.}
\label{fig:Average_SINR_vs_freq_diff_beta_Pskleft_right}
\end{figure}

\begin{figure}
\centering\includegraphics[width=0.72\linewidth,trim=0 0 0 0,clip=false]{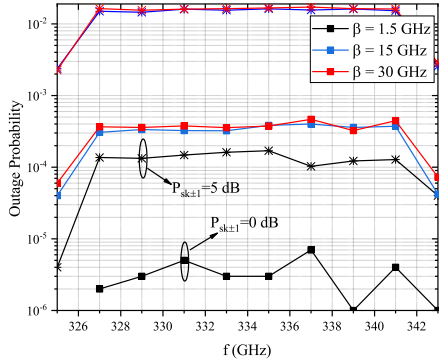}
\caption{OP vs $f_k$, for different $\beta$, and $P_{k\pm1}$, assuming $d=10\text{ }m$.}
\label{fig:OP_vs_freq_diff_Pskleft_right_diff_beta_fig1}
\end{figure}

Figs.~\ref{fig:Average_SINR_vs_freq_diff_beta_Pskleft_right} and~\ref{fig:OP_vs_freq_diff_Pskleft_right_diff_beta_fig1} depict the impact of LO PHN in the signal quality and system performance.
 From Fig.~\ref{fig:Average_SINR_vs_freq_diff_beta_Pskleft_right}, we observe that at each carrier frequency, the lower values of the LOs $3\text{-}\mathrm{dB}$ bandwidth and neighboring carriers transmission powers yield higher average SINR.  Fig.~\ref{fig:OP_vs_freq_diff_Pskleft_right_diff_beta_fig1} reveals that for $\beta>W_{ch}$, all the power of the neighboring $k\pm1$ carriers is treated as interference in carrier $k$. In other words, this result reveals that the LO should be selected in accordance with the carrier~bandwidth.

\begin{figure}
\centering\includegraphics[width=0.72\linewidth,trim=0 0 0 0,clip=false]{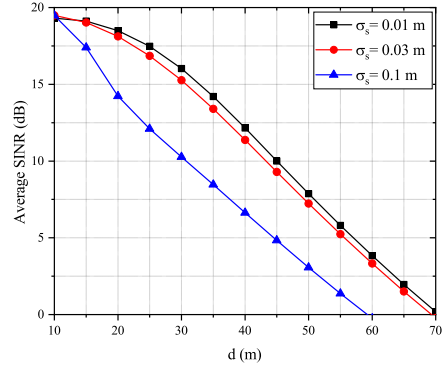}
\caption{Average SINR vs $d$, for different values of spatial jitter $\sigma_s,$ $P_{k\pm1}=5\text{ }dB$ and $\beta=1.5\text{ }GHz$.\\}
\label{fig:Avg_SINR_vs_d_diff_jitter}
\end{figure}

Fig.~\ref{fig:Avg_SINR_vs_d_diff_jitter} quantifies the impact of misalignment fading on the received signal quality. 
For a fixed level of misalignment fading, as the transmission distance increases, the path-loss increases; hence, the average SINR decreases. 
Furthermore, it is observed that for a given transmission distance, the increase of the spatial jitter leads to significant SINR degradation.  
This indicates the importance of taking into account the impact of misalignment fading, when evaluating the performance of a wireless THz~system.

\begin{figure}
\centering\includegraphics[width=0.72\linewidth,trim=0 0 0 0,clip=false]{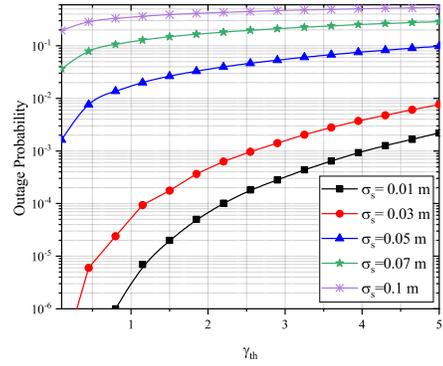}
\caption{OP vs $\gamma_{th}$ for different values of spatial jitter $\sigma_s$, $P_{k\pm1}=5\text{ }dB,$ $\beta=1.5\text{ }GHz$ and $d=10\text{ }m$.}
\label{fig:OP_vs_gthr_diff_jitter.eps}
\end{figure}

Fig~\ref{fig:OP_vs_gthr_diff_jitter.eps} presents the effect of misalignment fading to the wireless THz system outage performance. 
Moreover, for a given $\sigma_s$, as $\gamma_{th}$ decreases, the outage probability also decreases. This indicates that a countermeasure to the outage performance degradation, caused by misalignment fading, is the use of transmission schemes with lower spectral~efficiency.  

To sum up, our results revealed the frequency dependent behavior of the impact of PHN and provided useful insights for the selection of suitable LOs for wireless THz systems. Finally, the importance of taking into account the effect of misalignment fading, when quantifying the performance of these systems, was highlighted.

\vspace{-0.2 cm}
\bibliographystyle{IEEEtran}
\bibliography{References2}

\end{document}